\documentclass[10pt,letterpaper,compsoc,conference]{iiswc23}

\usepackage{cite}
\usepackage{amsmath,amssymb,amsfonts}
\usepackage{algorithmic}
\usepackage{graphicx}
\usepackage[dvipsnames]{xcolor}
\usepackage[final]{microtype}
\usepackage[italic]{mathastext}
\usepackage{libertine}
\usepackage[T1]{fontenc}
\usepackage{textcomp}
\usepackage[varqu,varl]{zi4}
\usepackage[all]{nowidow}
\usepackage[keeplastbox]{flushend}
\usepackage{soul}
\usepackage{listings}
\usepackage{xcolor}
\usepackage{todonotes}
\usepackage{multirow}
\usepackage[table,xcdraw]{xcolor}
\usepackage{authblk}

\definecolor{codegreen}{rgb}{0,0.6,0}
\definecolor{codegray}{rgb}{0.5,0.5,0.5}
\definecolor{codepurple}{rgb}{0.58,0,0.82}
\definecolor{backcolour}{rgb}{0.95,0.95,0.92}

\lstdefinestyle{mystyle}{
    backgroundcolor=\color{backcolour},   
    commentstyle=\color{codegreen},
    keywordstyle=\color{magenta},
    numberstyle=\tiny\color{codegray},
    stringstyle=\color{codepurple},
    basicstyle=\ttfamily\footnotesize,
    breakatwhitespace=false,         
    breaklines=true,                 
    captionpos=b,                    
    keepspaces=true,                 
    numbers=left,                    
    numbersep=5pt,                  
    showspaces=false,                
    showstringspaces=false,
    showtabs=false,                  
    tabsize=2
}

\begin{document}

%% EDIT TITLE BELOW

\title{The Non-Predictability of Mispredicted Branches using Timing Information}
%\title{Execution Latency does not Correlate with Hard to Predict Branch Outcome}

%% DO NOT EDIT THE FOLLOWING

%\renewcommand\Authsep{\qquad}
%\renewcommand\Authand{\qquad}
%\renewcommand\Authands{\qquad}

%% EDIT AUTHOR LIST BELOW

%\author{Author1 Name}
%\author{Author2 Name}
%\author{Author3 Name}
%\affil{Full Name of Awesome School}

%%% ALTERNATIVE FORMAT FOR MULTIPLE SCHOOLS:
%%% 
%\author[]{}
\author[1]{Ioannis Constantinou}
\author[2]{Arthur Perais}
\author[1]{Yiannakis Sazeides}
% \author[1]{Author4 Name}
\affil[1]{University of Cyprus }
\affil[2]{Univ. Grenoble Alpes, CNRS, Grenoble INP, TIMA, Grenoble, France}
\maketitle

% force page numbers
\thispagestyle{plain}
\pagestyle{plain}

%% EDIT YOUR PAPER'S CONTENTS BELOW
\begin{abstract}
Branch misprediction latency is one of the most important contributors to performance degradation and wasted energy consumption in a modern core. State-of-the-art predictors 
%such as TAGE or perceptron-based
generally perform very well but occasionally suffer from high Misprediction Per Kilo Instruction due to hard-to-predict branches.

In this work, we investigate if predicting branches using microarchitectural information, in addition to traditional branch history, can improve prediction accuracy.
%that one limitation of previous branch predictors is that they use only past branch history information available at prediction time. 
Our approach considers branch timing information (resolution cycle) 
%\hl{and on the prediction outcome of resolved branches (whether they are mispredicted or not)}. 
%These information is considered 
both for older branches in the Reorder Buffer (ROB) and recently committed, and for younger branches relative to the branch we re-predict. We propose Speculative Branch Resolution (SBR) in which, N cycles after a branch allocates in the ROB, various timing information is collected and used to re-predict.

%In this work, we attempt to show that one limitation of previous branch predictors is that they use only past branch history information available at prediction time. 
%In our approach we re-predict a branch instruction at a given cycle after it allocates in the Reorder Buffer (ROB) and use other information to . This allows to have more recent information to correlate on, and especially timing information of previous branches (e.g. resolution cycle, mispredicted/correct) but also of younger branches relative to the one we try to re-predict.

Using the gem5 simulator we implement and perform a limit-study of SBR
%(SBR) in which, N cycles after a branch allocates in the ROB, we collect various temporal features and re-predict the branch
using a TAGE-Like predictor. Our experiments show that the post-alloc timing information we used was not able to yield performance gains over an unbounded TAGE-SC. However, we find two hard to predict branches where timing information did provide an advantage and thoroughly analysed one of them to understand why. This finding suggests that predictors may benefit from specific microarchitectural information to increase accuracy on specific hard to predict branches and that overriding predictions in the backend may yet yield performance benefits, but that further research is needed to determine such information vectors.
\end{abstract}

\section{Introduction}

Despite its introduction 40 years ago by Smith~\cite{smith1981study}, dynamic branch prediction remains not completely solved and a critical performance feature.
%AP: 2/12 - Removing. determinant of the performance of high performance cores found in general purpose processors. 
%AP: 2/12 - Removing. Accurate branch prediction ensures the smooth delivery of instructions to the back-end of a core.
A clear indication of the significance of branch predictors is the large real estate they occupy in modern cores, which is reported to be tens of KBs~\cite{grayson2020evolution}. A branch prediction unit typically consists of a variety of predictors aimed at different types of branches, one of which predicts the direction of conditional branches (henceforth referred to as the branch predictor).

\begin{figure}[ht]
\centerline{\includegraphics[trim={0cm 0cm 0cm 0cm},clip,width=0.9\linewidth]{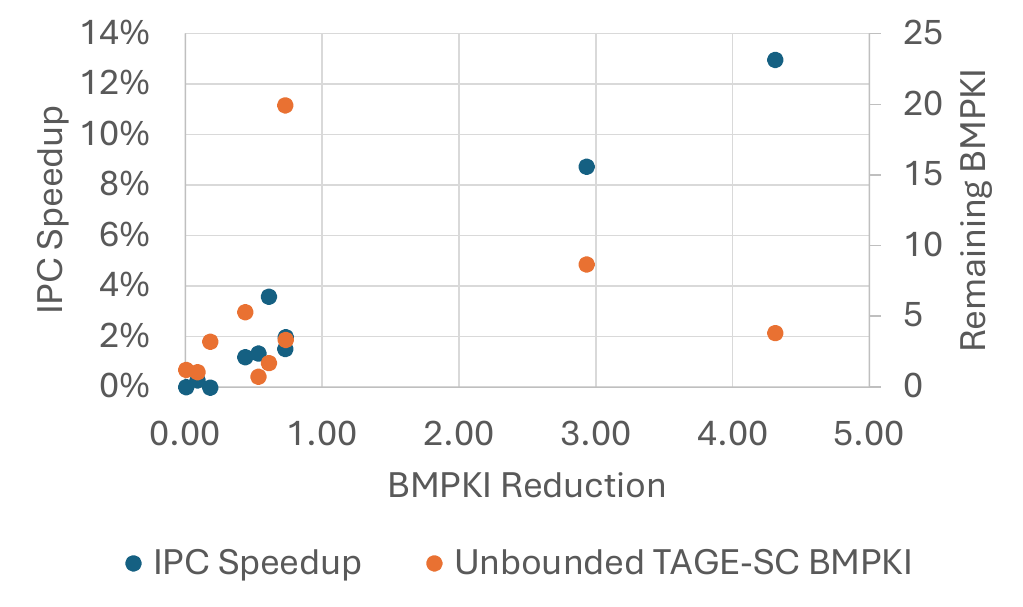}}
\caption{IPC Speedup and BMPKI Reduction of Unbounded TAGE-SC over a 64KB TAGE-SC (Secondary y-axis showing remaining BMPKI of Unbounded TAGE-SC for a specific BMPKI reduction compared to the baseline)}
\label{fig:huge_vs_normal_tage}
\end{figure}

\begin{figure}[ht]
\centerline{\includegraphics[trim={0cm 0cm 0cm 0cm},clip,width=1\linewidth]{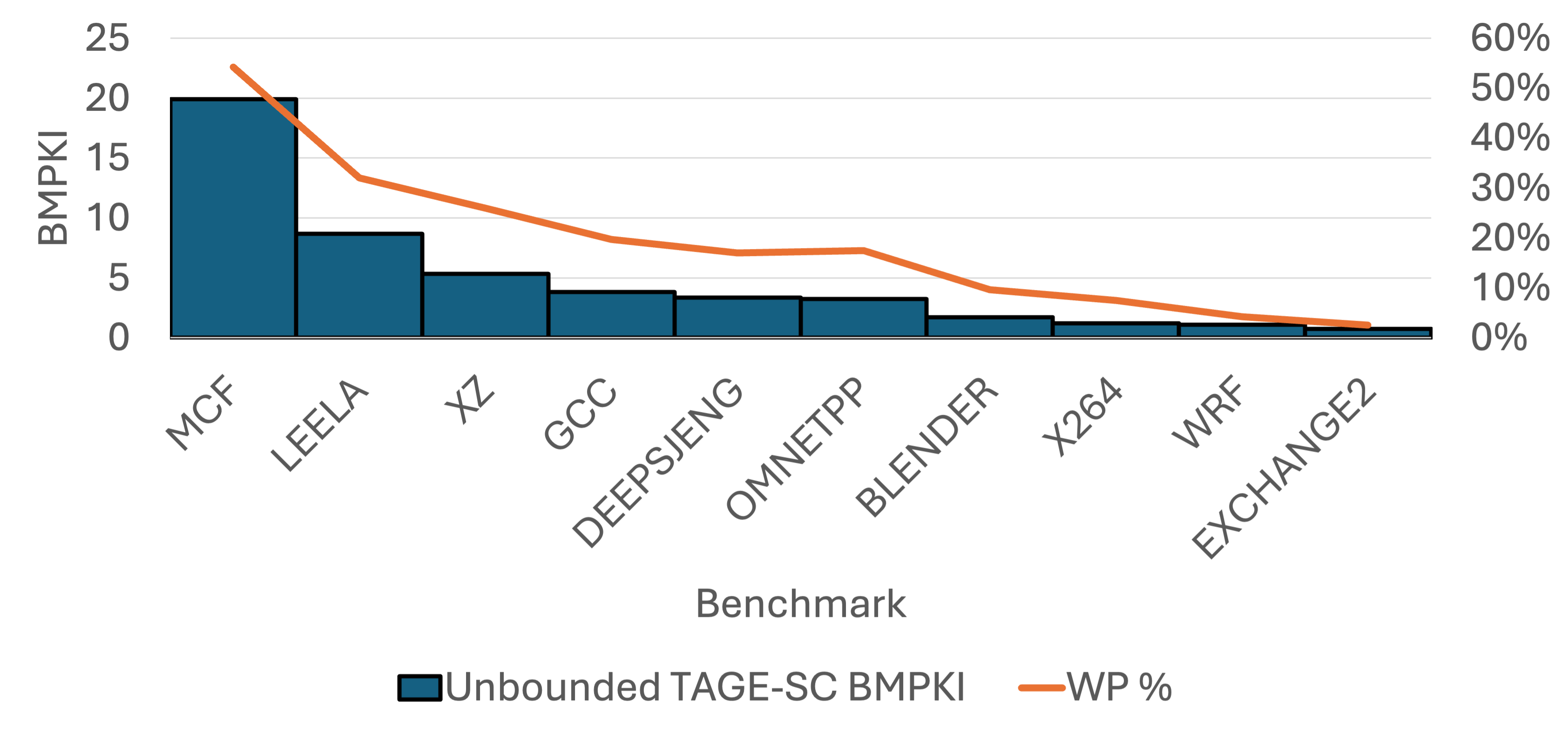}}
\caption{BMPKI of Unbounded TAGE-SC (primary y-axis) and \% of Wrong Path Instructions (secondary y-axis}
\label{fig:huge_tage_mpki}
\end{figure}

%AP: 2/12 - Removing
%\begin{figure}[ht]
%\centerline{\includegraphics[trim={0cm 0cm 0cm 0cm},clip,width=1\linewidth]{HUGE_TAGE_MPKI_WP.png}}
%\caption{BMPKI of Unbounded TAGE-SC (primary y-axis) and \% of Wrong Path Instructions (secondary y-axis}
%\label{fig:huge_tage_mpki}
%\end{figure}

Although state-of-the-art accuracy is high, Chauhan et al.~report that the performance of a Skylake-Like core would increase by 40\% with a perfect branch predictor~\cite{chauhan2020auto}, highlighting the significant potential for improvement. This is supported by Figure~\ref{fig:huge_vs_normal_tage} that depicts the (conditional) \textit{Branch Mispredictions Per Kilo Instructions} (BMPKI) reduction and \textit{Instruction Per Cycle} (IPC) speedup of an unbounded TAGE-SC~\cite{seznec2014tage} over a 64KB TAGE-SC on a set of SPEC2K17 rate benchmarks simulated on gem5 \cite{lowe2020gem5} (experimental methodology is presented in Section~\ref{exp_setup}). Clearly, there is a strong correlation between BMPKI reduction and performance. A potential solution to improve branch prediction accuracy is enhancing predictors table efficiency to reduce \textit{aliasing}, e.g., through larger tables.
%-- e.g., using 32M-entry TAGE tables without altering history lengths -- can significantly reduce BMPKI, as seen in Figure~\ref{fig:huge_vs_normal_tage}. 
However, even with effectively unbounded resources, many mispredicted branches remain (shown on the secondary y-axis in Figure~\ref{fig:huge_vs_normal_tage}), indicating the persistent challenge of Hard-To-Predict (H2P) branches.

High BMPKI is detrimental as it leads to a large fraction of wrong path instructions in the Reorder Budder (ROB). Figure~\ref{fig:huge_tage_mpki} highlights this relationship, showing that benchmarks with high BMPKI also exhibit a high percentage of wrong path instructions, with over 50\% in \textit{mcf}. Such wrong path instructions represent wasted computation and energy, though they may offer some prefetching benefits~\cite{pierce1996}. This highlights the significant opportunity to improve performance, particularly for workloads with complex control flow that defeats advanced branch predictors.

Prior work has proposed diverse approaches for dealing with H2P branches, including hardware-software co-designs for data-dependent branches~\cite{zilles2000,al2010exact,sheikh2012control,huawei2023} and hardware-only solutions like early misprediction detection using microarchitectural events that are more likely to occur in the wrong path (e.g., illegal opcode, nested mispredictions, etc.) ~\cite{armstrong2004wrong}. A recent study employed offline-trained convolutional neural neural networks to dynamically detect key features in the branch history to use for prediction~\cite{zangeneh2020branchnet}.

%H2P conditional branches have been the focus of many works that have identified many such branches to be data dependent and proposed hardware-software co-design approaches to predict them correctly~\cite{zilles2000,al2010exact,sheikh2012control,huawei2023}. One other recent work proposed a dynamic predictor that employs a convolutional neural network that is trained offline to detect online which features in the branch history to use for prediction~\cite{zangeneh2020branchnet}. A hardware-only approach explored by Armstrong et al.~\cite{armstrong2004wrong} aims to detect branch mispredictions earlier before they are resolved instead of improving the branch predictor accuracy. This approach relies on microarchitectural events that are more likely to occur on the wrong path (e.g., illegal opcode, misprediction under misprediction etc.) to recognize that execution is likely in the wrong-path and resolve faster branches that are predicted to be mispredicted. 

In this work, we first note that virtually all state-of-the-art predictors rely on \textit{architectural} information to predict branches. A typical example is the global branch history, which represents the directions of past conditional branches and does not depend on the microarchitecture, but only on the program being executed. 
%AP: 2/12 - Removing. Indeed, even if the representation of this history within the microarchitecture is speculative, if a branch predicted using a particular history retires, then this means the speculative global branch history used to predict it was the correct one, that is, it eventually became part of the architectural state. 
As a result, we explore the potential for exploiting correlation in \textit{microarchitectural} information to predict branches. While there are many microarchitectural features that can be explored (e.g., cache misses, ROB occupancy, etc.), we focus on \textit{timing}, i.e., the fact that at the moment we predict, some events have taken a given amount of \textit{time}. The rationale behind this, is that timing can encode many specific events of the microarchitecture without having to explicitly watch each individual event. Another interesting aspect of this approach is that because the microarchitecture has a speculative window of "future" instructions, it is even possible to use the post-fetch timing behavior of younger instructions to (re-)predict older but not yet resolved branches, naturally leading to an overriding prediction approach. Using "future" -- yet specific -- information was suggested as a way to improve early resolution coverage by Armstrong et al.~\cite{armstrong2004wrong}.

In summary, this paper first contributes a real-world example presenting why microarchitectural information and in particular timing can serve as an information vector for prediction. Second it proposes a microarchitectural flow for overriding the fetch predictor using microarchitectural information, \textit{Speculative Branch Resolution}, or SBR. Third, it studies different combinations of microarchitectural information vectors to estimate the potential for SBR to provide meaningful MPKI gains. Although the results are generally negative -- the microarchitectural information vectors used in our studies are not able to beat an impractically large TAGE-SC predictor -- we provide a detailed analysis as to why and suggest cases where timing information may still have value.

\section{Related Work}

% mentione din intro no need to repeat
%Dynamic branch prediction is a significant contributor to processor performance and was initially proposed by Smith \cite{smith1981study}. Since then, many refinements have been proposed. 

%We review some of the existing work in dynamic branch prediction.
\subsection{Mechanisms that Leverage Architectural Information}

Most predictors rely on tying a program context to a given direction prediction. For instance, many predictors rely on the global branch history ($n$-bit vector with the direction of the $n$ most recent conditional branches). These contexts are \textit{architectural} in the sense that they depend only on the program semantics, and not on the hardware the program is being ran on. 
Many algorithms leveraging this type of information were proposed \cite{yeh1991two,eden1998yags,mcfarling1993combining,seznec2002design}, but this section only focuses on state-of-the-art predictors.

\paragraph{TAGE}
This predictor \cite{seznec2006case} combines the global branch history and the path history (lower bits of past branch PCs)
%concatenated with most recent PCs towards LSB) 
to index various tagged tables containing prediction counters. Each table is accessed with a different global history length, and the lengths follow a geometric series. This allows to capture correlation between distant branches, which are typically rare, while dedicating most of the storage to correlation between close branches. Improvements over a fine-tuned TAGE are generally achieved by combining it with "add on" predictors. One example is L-TAGE \cite{seznec2007ltage} where a loop predictor tries to identify regular loops with long loop counts. Another variant is TAGE-SC-L \cite{seznec2014tage} that, in addition to the loop predictor, features a Statistical Corrector component which tries to find branches that are statistically biased and that TAGE does not always predict correctly. Another "add on" is the Inner Most Loop Iteration counter, or IMLI \cite{seznec2015inner}. The IMLI counter is added in the statistical corrector and helps predicting branches that are encapsulated in the innermost loop body of multidimensional loops, by correlating the loop count with the encapsulated branch outcome.

\paragraph{Perceptron}
The original perceptron-based predictor \cite{jimenez2001dynamic}, and its refinements \cite{jimenez2003fast,tarjan2005merging,jimenez2016multiperspective}, use a different method of predicting branches based on neural networks. A single layer of perceptrons computes a prediction using a linear function that takes architectural information and per-PC weights as input to produce a prediction.

\paragraph{BranchNet}
This predictor \cite{zangeneh2020branchnet} uses a Convolutionnal Neural Network (CNN) architecture to predict branch directions. BranchNet is trained offline using program traces. At runtime, BranchNet predicts delinquent H2P branches, while the remaining branches are predicted by TAGE-SC-L. This hybrid approach can reduce the BMPKI of SPEC2017 benchmarks by 7.6\% (and up to 15.7\%) when compared to a standalone unlimited MTAGE-SC \cite{seznec2016exploring}, and performs well because the CNN scheme can detect correlations in noisy and very long global histories without the need for exponentially large storage as the history grows.

\paragraph{Overriding Predictors}

Ideally, branch predictors should provide the predicted direction of a branch in the cycle the branch is fetched, so that subsequent instructions can be fetched in the next cycle. Unfortunately, to achieve high accuracy, predictors use large tables and cannot meet such timing requirement. As a result, branch prediction units can combine a small but fast reasonably accurate predictor with a large but slow and very accurate \textit{overriding} predictor \cite{jimenez2000impact,loh2006revisiting,seznec2002design}. Although the latter is slower and incurs an override penalty, that penalty is much lower than the full branch misprediction penalty. Our approach can be seen as having an overriding predictor in the backend.

\subsection{Mechanisms that Leverage Microarchitectural Information}

%The main prop microarchitectural information, that is, information that does not necessarily match program semantics but is a consequence of how instructions flow in the pipeline during speculative execution.

\paragraph{Prophet/Critic Hybrid Branch Prediction}

The prophet/critic hybrid branch predictor \cite{falcon2004prophet} has two predictors. The prophet is a classical predictor that correlates on the global branch history. The critic, on the other hand, re-predicts branches later in the pipeline, using information of "future" branches, that is, from predictions made by the prophet that correspond to branches that are younger than the one the critic is re-predicting. As a result, prophet/critic makes use of microarchitectural information because if critic overrides the prophet, then the predictions of the "future" branches used by the critic i) May not correspond to future correct control flow and ii) will disappear from the pipeline, despite having been used to generate a prediction.

\paragraph{Wrong Path Events}

Armstrong et al.~observe that rare events such as memory related exceptions (permission faults, unaligned accesses not supported in an ISA, etc.) can indicate that the pipeline is on the wrong path \cite{armstrong2004wrong}. As a result, branch mispredictions can be resolved early if the proposed heuristic deems that a "problematic" event taking place does indeed indicate the control flow is incorrect. Unfortunately, "Wrong Path Events" (WPE) are rare and thus not many branches can be resolved early in this fashion. Our work proposes to find more frequent WPE to correlate on, namely the state of younger branches. 

%Additionally, we investigate the benefit of learning correlations based on microarchitectural information -- such as branch timing -- instead of architectural information.

%AP: 2/12 - Removing which is is one of our paper aims to investigate.

%\todo[inline]{Need to elaborate more and add more related work? Done?}

%\todo[inline]{Sota branch prediction (TAGE, SC, IMLI, perceptron, etc). "Exotic things": Prophet/Critic, WP event to indicate mispredictions, branchnet? EXACT predictor?}

\section{Motivation for Timing Information Vector} \label{motivation}

%This section introduces a motivating example where timing information can be helpful to provide more correct predictions.

%Using toplev\cite{toplev} to perform top-down analysis and Linux perf tool to collect the BMPKI and the SPEC2017 rate \cite{bucek2018spec} CPU benchmarks we show the impact branch mispredictions have on modern hardware. Fig.~\ref{fig:topdown} illustrates the fraction of cycles during which CPU was stalled due to branch mispredictions (blue column). It also reports the BMPKI of each benchmark (red line). We can observe that up to 30\% of the time is spent on wrong path execution, with this ratio correlating to the workload BMPKI. This demonstrates that control flow speculation remains an important contributor to performance and motivates the need for finding new ways to predict hard-to-predict, or H2P, branches.

% @Ioanni: Don't bother with trimming the pdfs in excel or inkscape
% this will do the trimming in latex directly (parameter of % includegraphics command)
% clip,trim={<left> <lower> <right> <upper>}

%\begin{figure}
%\centerline{\includegraphics[trim={0cm 0cm 0cm 0cm},clip,width=\linewidth]{Potential_Skylake_all.pdf}}
%\caption{Cycles attributed to "Bad Speculation" (blue columns) and BMPKI (red line). Skylake core}
%\label{fig:topdown}
%\end{figure}

As previously mentioned, even with unbounded tables, state-of-the-art branch predictors are unable to predict correctly many branches and, as a result, alternative prediction techniques should be devised.
%Those techniques should not rely on huge predictors that use very large tables to mitigate the aliasing stemming from using very large histories. 

In this work, we adopt the suggestion by Armstrong et al.~and enrich the type of information used to learn the behaviour of H2P branches~\cite{armstrong2004wrong}. More specifically, instead of using architectural information only (e.g., direction of past branches, address of past branches, etc.) as most of the existing predictors do, we also consider microarchitectural information such as branch timing  information.
%and whether a branch is predicted correctly or not. In this Section we focus specifically on timing information of %both older and younger -- i wish we had a poster child that had younger --- i do not mention older younger because our discussion shows only older 
%branch instructions.

\subsection{A Culprit} To illustrate why timing information can be used for prediction, and guide our selection of what timing information to use we analyze a H2P branch in the \textit{mcf} workload of SPEC2017 CPU, which is a notoriously hard workload for control flow speculation and and was one of the worst performers in Figure \ref{fig:huge_tage_mpki} (exact BMPKI numbers are provided in Table~\ref{tab:h2p} in Section \ref{exp_setup}).

Listing \ref{list:mcf_code} is an extract of the source code of the \textit{spec\_qsort} function, which contains six H2P branches with mispredict rates ranging from 11.1\% to 31.3\%. 
%\todo{Any chance you know how often? Like how many calls in the simpoint of %interest maybe?}
The \textit{quick\_sort} algorithm, given an unsorted table \textit{t} of length \textit{n}, determines a \textit{pivot} value in the table that is used to move the values that are lower or equal to (resp.~greater than) the pivot in the lower (resp.~higher) part of the table. This process is recursively repeated on the derived table partitions. 
%AP: 2/12 - Removing \textit{mcf} uses \textit{spec\_qsort} to sort the arcs of a graph, and 
The Listing corresponds to the part of the algorithm that decides, according to the size of a partition, how to select the pivot:

%\textcolor{orange}{AP: I put the Figure backremoved the Figure, fix this? Or add the figure back. Ioannis Answer: I can fix this but maybe we can add the Figure back if we make the space? Figure was showing BMPKI and WP insts \%. } Figure \ref{fig:huge_tage_mpki}. 

\noindent
For short partitions, $n<7$ (lines 4-9), it uses insertion sort to sort a partition and end the recursion.\\
For $n>=7$, it partitions the table and recurses using sampling for pivot selection as follows:
\begin{itemize}
    \item For $n=7$ (Line 10), use the value in $t[n/2]$ (value in the middle of the unsorted partition),
    \item For $n<=40$ (Line 12-21), use the median of $t[0]$, $t[n/2]$, $t[n-1]$,
    \item For $n>40$ (Line 15-20), use a more complex selection from 9 values
\end{itemize}

Branch A (Line 14), which checks if $ n >40$, is the H2P branch we focus on and its mispredict rate is 18.6\%. Note that \textit{spec\_qsort} includes a recursive call to itself with its lower partition as parameter and a goto statement to iterate over part of the \textit{spec\_qsort} function when it returns from the recursive call to process the upper partition (thus avoiding the overhead of some recursive calls). As result, each function invocation may execute branch A multiple times with a recursive call between different executions of branch A. 

%\textcolor{blue}{Can we cheat and make the listing smaller? Text smaller?}

\begin{lstlisting}[language=C, caption=Source code of spec\_qsort in mcf,label={list:mcf_code},basicstyle=\fontsize{7}{7}\selectfont\ttfamily]
void spec_qsort(void *a, size_t n, size_t es, cmp_t *cmp){
... // Skipping some lines of code ...
loop: 
    if ( n < 7) {
        for(pm = (char *)a + es; pm < (char *)a + n * es; pm = +es)
           for (pl = pm;pl > (char *)a & cmp(pl - es, pl) > 0;pl=- es)
                   swap(pl, pl - es);
        return;
    }
    pm = (char *)a + (n / 2) * es;
    if (n > 7) {
        pl = (char *)a;
        pn = (char *)a + (n - 1) * es;
        if (n > 40) {  // Branch A
            d = (n / 8) * es;
            pl = med3(pl, pl + d, pl + 2 * d,cmp);
            pm = med3(pm -d, pm, pm + d, cmp);
            pn = med3(pn - 2 * d, pn - d, pn, cmp)
        }
        pm = med3(pl, pm, pn, cmp);
    }
    ... // Skipping some lines of code that 
    ... // perform the partitioning and update
    ... // local variables pa, pb, pc, pd, pn...
    if ((r = pb - pa) > es)
        spec_qsort(a, r / es, es, cmp);
    if ((r = pd - pc) > es) {  // Branch B
    /* Iterate rather than recurse to save stack space */
        a= pn- r;
        n = r / es;
        goto loop;
    }
}
\end{lstlisting}

%\todo[inline]{fix: the motivation data seem a bit redudant with figure 1 if you will kep them need to fid a way to highlight both in the intro}.

%\todo[inline]{fix: impractcial -fix it everywhere yo unbounded and explain in methodolgoy what is meant by unbounded}

%\todo[inline]{fix: it is good also to check if this behavior holds for the 64KB - the writeup needs cleanup why you assume B is older than A, it is not assume it can happen and explain how - i think you need to explain better - the story here -- to many assumptions - we have analyzed the size of the records... also isnt it good enough to know that when this branch is delayed its behavior changes... show its direction and mispredict rate as a fiunction resolution time -- i think this is a very simple and nice behavior all you need to heck is this branch time unresolved... this swhat you need to show -- and given you establish this then you try to explot it in general this type of behavior interplay of mispredicts/branch direction with timing of branches younger or odler}

%\todo[inline]{fix: keep it simple do not mix online and offline}

\begin{figure}
\centerline{\includegraphics[trim={0cm 0cm 0cm 0cm},clip,width=\linewidth]{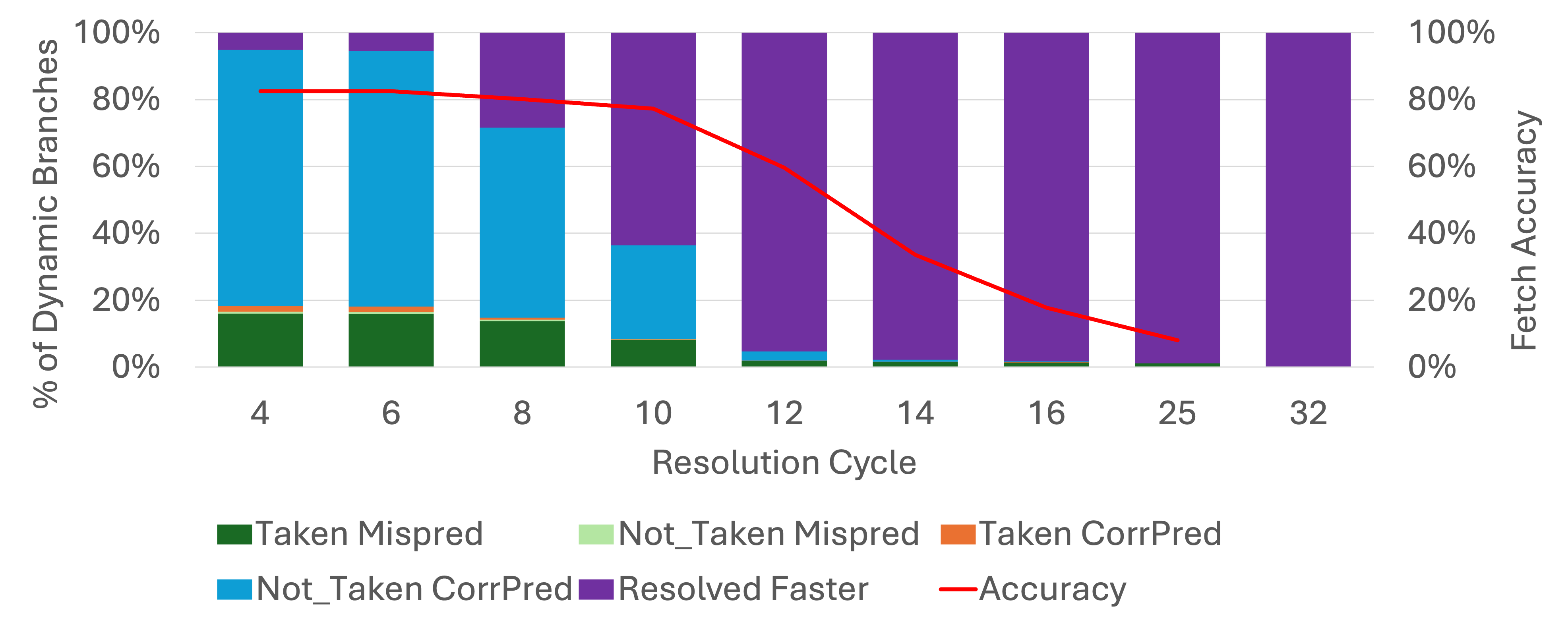}}
\caption{Taken and Not-Taken Correct Predictions and Mispredictions of Branch A as a function of resolution time (normalized to the total executions of this branch)}
\label{fig:mcf_mr_rates}
\end{figure}

\subsection{Analysis} 
%\textcolor{orange}{Do we need to mention all the categories in text ? Should we just go to conclusion? Ioannis Answer: Not sure which categories you talk about here. AP: I mean "unresolved not taken and predicted correctly" then "unresolved and taken and predicted correctly" etc. Written like this, it takes 4/5 lines, I am wondering if we can jump to conclusion and let the Figure show the categories.}
Figure~\ref{fig:mcf_mr_rates} reports some statistical properties of committed instances of branch A, as a function of the number of cycles that elapsed since the branch was allocated in the ROB (x-axis) normalized to the total number of committed instances of branch A. By cycle four: 5.2\% of instances have resolved, 76.8\% are unresolved and will eventually be not-taken and predicted correctly, 1.7\% are unresolved and will be taken and predicted correctly, 0.5\% are unresolved and will be not-taken and mispredicted and finally, 16.0\% are unresolved and will be not-taken and mispredicted. What we observe is that the \textit{if} statement is mostly false (corresponding to not-taken branch A) and when it is true (A taken), it is almost always mispredicted. As the number of elapsed cycles after allocation in the ROB -- \textit{post-alloc} -- increases the number of branches that remain unresolved decreases from 95.4\% in cycle 4 to 71.6\% in cycle 8, 2.3\% in cycle 14, 1.7\% in cycle 16 and 1.2\% in cycle 25 (a handful remain unresolved past cycle 32). The key observation here is that the longer an instance of A remains unresolved, the more likely it is to be mispredicted! The Figure shows that the accuracy (secondary y-axis) decreases from 82.6\% in cycle 4, to 80.1\% in cycle 8, to 33.6\% in cycle 14 and 17.8\% and 7.8\% in cycles 16 and 25, i.e., the large majority of instances that remain unresolved for 14 or more cycles are mispredicted.

Further investigation first revealed that the corresponding \textit{if} statement line 14 is mostly false (i.e., $n$ is generally less than or equal to 40) because, assuming the \textit{spec\_qsort} partitions are perfectly balanced, then checking for a given size of partition $p$, as branch A tests, will converge to be true $1 \over p$ times.\footnote{Assuming a perfect binary tree with each node, except the root, representing a quick\_sort partition of a table with size 2$^n$, then there are 2$^{n+1}$-2 partitions in total, each with half the size of its parent node. In such tree, the fraction of nodes with partition size greater than or equal to 2$^k$ is given by 1 - 2$^{n}$/2$^{k}$ \ * (2$^{k+1}-2$)/(2$^{n+1}-2$) which simplifies to 1/2$^k$ assuming 2$^{n+1}$ is much larger than $2$.} For instance for $n=41$ and table size greater than 1000 we expect around 2.5\% of the instances of branch A to be true. The reason that the branch is true around 18\% is because the \textit{spec\_qsort} stops partitioning when the partition size is $< 7$ and uses insertion sort instead. Next, we focused on why this \textit{if} is mostly false and branch A is mispredicted when it takes more than 14 cycles to resolve.

We observe that on occasions, when branch A is executed following a return from a recursive call, its execution is slower (Figure~\ref{fig:mcf_mr_rates}) for values of $n$ greater than 40. This is due to the eviction of the local variables that the branch tests from the L1D. This is more likely to occur for larger values of $n$ due to the amount of data being accessed: each arc records in the table to be sorted is 72 bytes, and the deeper recursion increases the number of allocated stack frames. Furthermore, when we are in a \textit{spec\_qsort} invocation with value of $n$ greater than 40 we expect for branch A to be executed \textit{at least} seven times (this number depends on how well balanced the partitions are as we recurse through the data) until we return back to this function invocation, and all of these seven tests outcomes to be false, because at least seven resulting partitions will have length less than 40. This means that the predictor is trained multiple times with the information that branch A's outcome ($n>40$) is false and as a result, when we return to the invocation with value of $n$ equal to 40, the branch is usually mispredicted.

%as the TAGE predictor used in this study cannot de-alias history patterns. 

%\begin{figure}
%\centerline{\includegraphics[trim={0cm 0cm 0cm 0cm},clip,width=\linewidth]{vectors_scatter_mcf.png}}
%\caption{Scatter plot of vectors encoded with 1 and 0 based on resolution cycles for mcf H2P branch (Length is 10 branches)}
%\label{fig:vectors_scatter_mcf}
%\end{figure}

The discussion so far established that the number of cycles branch A remains unresolved can help detect whether it is mispredicted and, therefore, that timing information of microarchitectural events can serve as a source of information for branch prediction. However, at least for this example, the potential of this approach is rather limited since only a fraction of the mispredicted instances of A remain unresolved after cycle 12 (see Figure~\ref{fig:mcf_mr_rates}), and for earlier cycles, most instances are predicted correctly. However, one can also correlate on the timing behavior of other branches. Specifically, in this example, between the time we return from a recursive call and the execution of A, six branches are often executed, with each of these branches testing information contained in local variables. As a result, correlating on the timing behavior of some or all of these branches can actually serve us better because a possible mispredict by any of these branches may hide the cache delay of the inputs of branch A. In fact, we found that whenever the \textit{if} statement corresponding to branch B line 27 is true (execute the goto), and its resolution required more than 20 cycles (roughly the latency of the L2 cache), branch A is mispredicted 82\% of the time. Conversely, when the \textit{if} statement is true but B resolution is less than 20 cycles, branch A is only mispredicted 15\% of the time.

\subsection{Implications and Key Parameters}

The previous example established that the timing of a branch -- as far as how many cycles it remains unresolved post allocation in the ROB -- can serve as an information vector for prediction for itself and for other branches. This vector can then be used to resolve branches after a number of cycles to reduce their misprediction penalty.

%AP: 3/12 - Removing
%, rather than wait for them to execute to resteer the pipeline. 

A key parameter of a timing based prediction approach is the number of cycles to wait to re-predict a branch. Beyond its accuracy and coverage implications, the re-predict cycle is important for the performance gain or penalty of such a scheme. We illustrate this with an example: when branch A remains unresolved for 16 cycles after it is allocated, if resolved speculatively it can save at least 10 cycles for 1.4\% of the branches ((100\%-17.8\%) x 1.7\% = 1.4\%, with 17.8\% corresponding to the prediction rate of branches unresolved in cycle 16 and 1.7\% the fraction of branches that remain unresolved after 16 cycles). These savings are possible because as we see in Figure~\ref{fig:mcf_mr_rates}, almost all branches that are unresolved in cycle 16 remain unresolved until cycle 25. This benefit comes at the expense of increasing the delay for 0.3\% of the branches (17.8\% x 1.7\% = 0.3\%), who were initially predicted correctly but incorrectly resolved speculatively. Speculatively resolving branches early increases the opportunity for SBR as more branches are likely to remain unresolved but the accuracy of early resolution may be lower, hence rendering the overall approach detrimental. In the same vein, waiting to resolve speculatively later can help increase accuracy but at the expense of lower opportunity (as more branches are likely to be resolved already).

\section{Speculative Branch Resolution} \label{sbr}

%AP: Removed for space and redundancy
%Our goal is to address the challenges posed by H2P branches by leveraging timing information, which in this context refers to the post-alloc resolution latency. 
One key issue with using timing information
%, and especially how much time instruction take to execute, 
is that it is usually not available at prediction time for many branches.
%AP: Making space
%Indeed, recently fetched branches are often still on their way to the backend and have not resolved yet, while younger branches have not even been fetched.
As a result, our proposed technique, \textit{Speculative Branch Resolution} or \textit{SBR}, is an overriding scheme that re-predicts branches in the backend to be able to leverage more rich timing information.

%In this section, we dive into the specifics of SBR, starting with an explanation of how we compose our timing information vector. Then, we detail how and when we utilize this information to re-predict H2P branches. 

%The emphasis of the exposition is on timing information vector but other microarchitectural information can be incorporated similarly (we discuss the various types of microarchitectural information we investigated in the Section~\ref{}).

\subsection{Timing Information Vector}

SBR repredicts a branch instance using timing information of other branch instances. At a given time, the \textit{timing} of a branch instance corresponds to the number of cycles it spent in the ROB unresolved (so far). Depending at which stage and when SBR repredicts a branch, the timing information that is available for other branches varies.

First, at fetch, we can obtain the timing information of retired branches as well as older branches that are already in the ROB, that is, a subset of older branches. In this case, the timing information of branches older than the fetched branch that have not been added to the ROB is not available. 
Another point is just before a branch allocates an entry in the ROB, which we term \textit{pre-alloc}. At this point, we can obtain the timing information of all the older branches in the ROB, as well as of retired branches. However, there is no timing information for branches that are younger as those branches have not yet reached the ROB. A last option is a few cycles after entering the ROB, which we term \textit{post-alloc}. In this case, we have timing information about i) Retired branches ii) Older branches still in the ROB and iii) Younger branches that entered the ROB. Assuming the re-predicted branch was initially mispredicted, waiting longer to re-predict reduces the number of cycles saved,  but enriches the available timing information. In this work, 
%we focus on the potential of SBR and, therefore, we 
SBR re-predicts post-alloc, as this provides the richest set of information to the overriding predictor. We define the number of cycles we wait before re-predicting a branch after it entered the ROB as the \textit{re-predict cycle}.

The \textit{Timing Information Vector}, or TIV, used by the SBR predictor 
%to provide predictions 
encodes the timing behavior of branches. A single bit of this information vector encodes whether a branch instance resolved fast or slow based on a \textit{timing threshold}. This does not distinguish between a branch being resolved or unresolved, although a resolved branch will never change category as time passes, while an unresolved branch may switch from fast to slow as cycles advance. The TIV is built from the concatenation of three sub-vectors:

%\textbf{TODO: Couple of example vectors maybe? must show info vectors - is there still inconsistency or this is fixed?}

%\todo[inline]{Add a section to explain in detail what Information vectors can be available}

%10/12 Fixed
%\textcolor{orange}{AP: This is not super clear because paragraph above says it can only be fast or slow and we don't distinguish resolved vs not, Ioannis Answer: We denote which of the Vectors can have resolved and unresolved branches and which cannot, but I get that it is confusing. Should I remove this sentence from OTIV and YTIV. AP: I get it, it might be a bit confusing but that's ok}

\textbf{Commit Timing Information Vector or CTIV} is a vector that contains timing information regarding branches that have retired, hence have resolved. Each entry in CTIV denotes whether a branch resolved fast or slow.

\textbf{Older Timing Information Vector or OTIV} is a vector that contains timing information about branches that reside in the Reorder Buffer (ROB) and are older relative to the branch being re-predicted. Those can either be resolved or unresolved.

\textbf{Younger Timing Information Vector or YTIV} is a vector that contains timing information of branches that are in the ROB and are younger relative to the branch being re-predicted. Those can either be resolved or unresolved.

%\textcolor{blue}{AP: "Threshold" has not been introduced here has it? Answer: It is introduced in the paragraph starting "The timing information..."}

Our approach uses branch resolution timing as a proxy for various potential sources of delays (e.g., TLB/cache misses) because such events are reflected in resolution delays. This captures the combined impact of all the delay inducing events without explicitly modeling each individually, enabling us to evaluate the potential upper bounds of leveraging aggregate timing information for re-prediction. 
Future work can investigate the benefits of isolating the source of delays in the TIV.

\begin{figure}
\centerline{\includegraphics[width=0.9\linewidth]{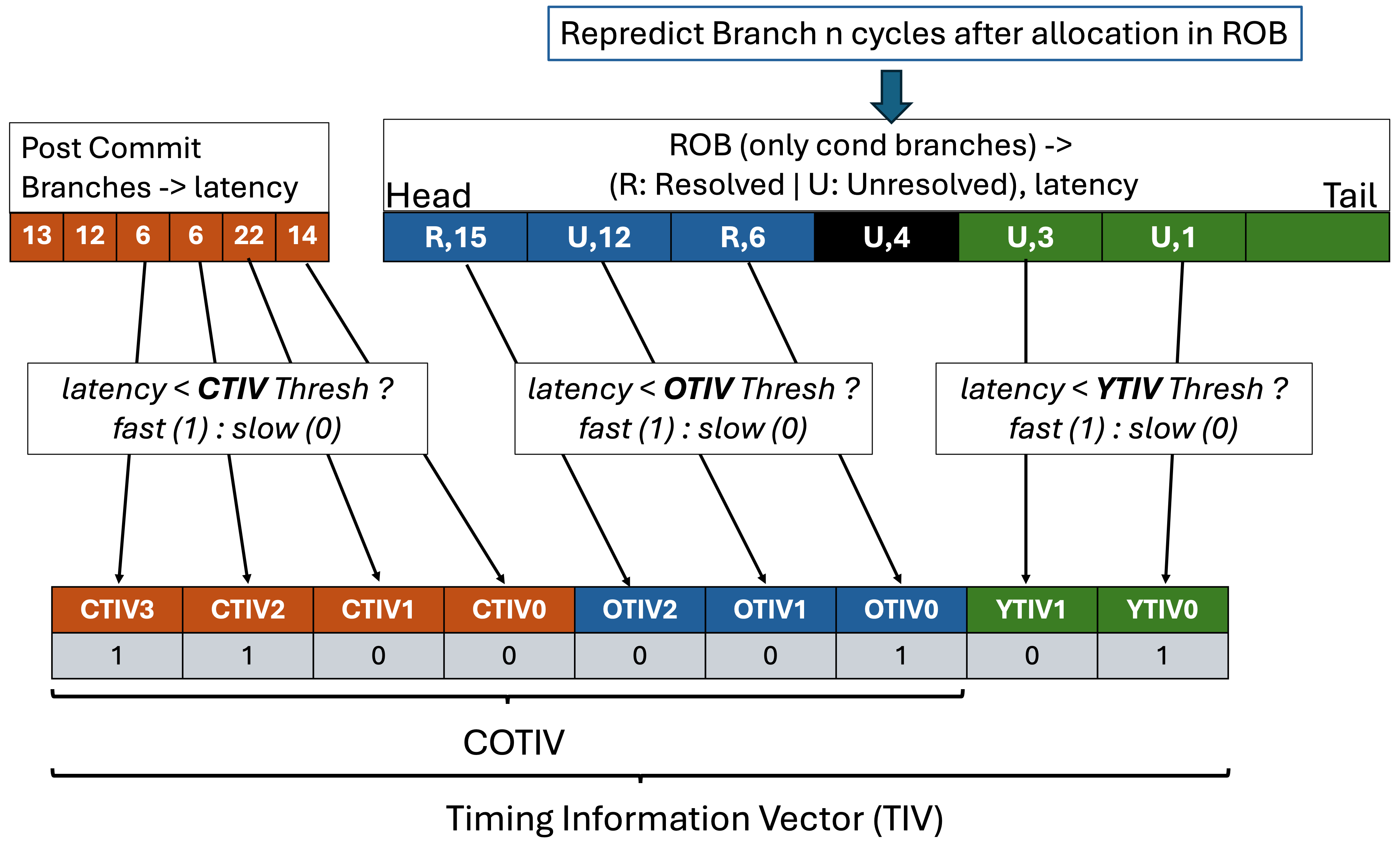}}
\caption{Example of how the Timing Information Vector is constructed (Re-predict cycle: 4, CTIV Threshold: 8, OTIV Threshold: 8, YTIV Threshold: 2)}
\label{fig:vector_example}
\end{figure}

Figure~\ref{fig:vector_example} shows an example of how a vector is constructed out of the ROB and the retired branch instructions, assuming re-predict cycle 4, CTIV and OTIV timing threshold of 8 and YTIV threshold of 2. The total size of the TIV for this example is assumed to be 9. The branch being re-predicted is highlighted in black. The younger branches are in green, the older but not yet committed are in blue, and the recently retired ones are in orange. We first check the number of cycles elapsed since allocation of younger branches, and determine whether they are fast or slow based on the YTIV threshold. For instance, although it is not resolved yet, the youngest (rightmost) one is considered fast because it has spent fewer cycles in the ROB unresolved than the YTIV threshold. Continuing, we go through all the branches in the ROB, except the re-predicted branch, and we set the bits in the TIV. If there are not enough branches in the ROB to complete the TIV, we go through the retired branches. For this example, we use 4 retired branches to fill the remaining slots in the TIV. 

A subtlety of the TIV construction algorithm is that assuming it has length of $n$, we can always build it using information from the $n$ most recent older branches by combining the OTIV and the CTIV. However, depending on timing, there may be an arbitrary number of younger instructions in the ROB. In practice, having more younger branches pushes out some bits from CTIV (and potentially OTIV), and makes two different TIV not directly comparable as the bit position that delineates older information from younger information, relative to the branch being re-predicted, varies. In addition, if the YTIV uses a different threshold, a branch may be considered slow in the YTIV, but fast when it moves to the CTIV or OTIV. Since the TIV is always of length $n$, this TIV construction algorithm "inconsistency" can help to encode how many elements of the TIV are from the YTIV vs. the CTIV/OTIV.
%AP: 10/12 Done
%\textcolor{orange}{AP: Did not follow, we meaning we could keep the info and use it? We don't right? Ioannis Answer: No we do not keep any extra info, I think we say by construction the TIV is "n" bits and when YTIV is inserted vs when it is not inserted TIV is inconsistent and SBR could capture this difference and quantify the number of YTIV inserted, not sure if I made it more clear or confusing}. 
We also observe that inserting the direction provided by the fetch predictor for the branch being re-predicted into the TIV is important to achieve better accuracy. Overall, the TIV vector combines the information in the following order: the commit vector, then the older vector, then the younger and finally the fetch prediction.

%Hence, the inclusion of YTIV may introduce inconsistency in history bit position. However, we \textit{I do not like "believe". Do we have any quantitative or qualitative argument here?}{believe} it enriches the timing information vector by providing "future" branch behaviour. 

\subsection{SBR Mechanism}

Speculative Branch Resolution (SBR) re-predicts H2P branches post-allocation but before resolution, functioning as an overriding predictor \cite{jimenez2000impact,loh2006revisiting,seznec2002design}.  %Although \textcolor{orange}{Loh argues that overriding predictors should override before renaming takes place, to avoid the need for restoring the rename map table \cite{loh2006revisiting}, SBR re-predicts in the backend to maximize the available timing information. Consequently, when SBR disagrees with the fetch re-predicted branches are treated as mispredictions, and flush the pipeline and repair the register mappings. This increases the cost of overriding, but is no more complex than handling a mispredicted branch in the specific microarchitecture in which SBR is added. }
The fetch predictor and the SBR overriding predictor we use are TAGE based and as a result, we differentiate between them as \textit{Fetch TAGE} predictor, and the \textit{SBR TAGE} predictor.

\begin{figure}
\centerline{\includegraphics[width=0.8\linewidth]{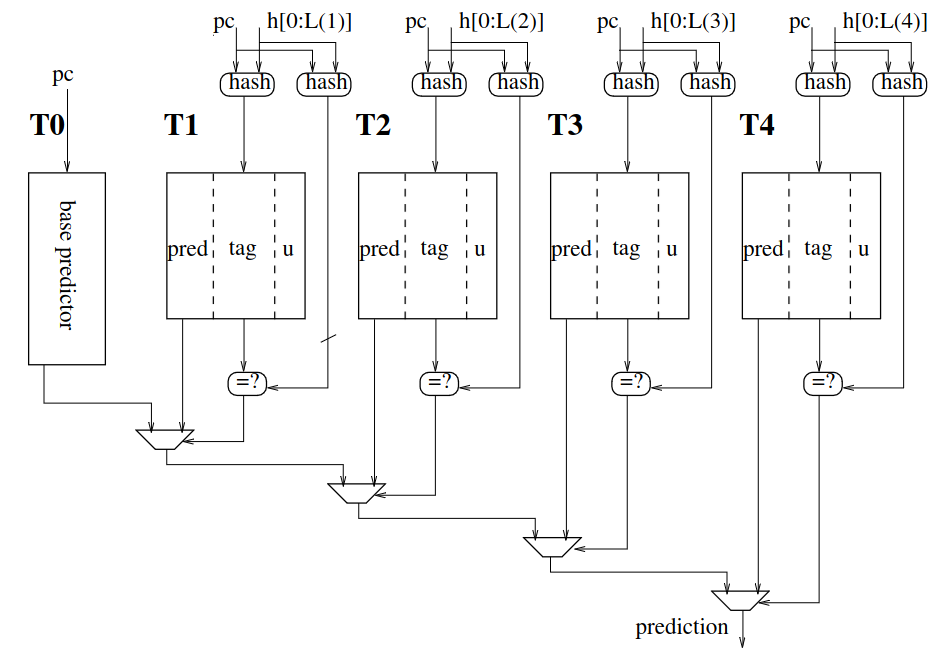}}
\caption{5-component TAGE predictor synopsis: a base predictor is backed with everal
tagged predictor components indexed with increasing history lengths. From Seznec and Michaud \cite{seznec2006case}.}
\label{fig:tage}
\end{figure}

\paragraph{TAGE Predictor Primer} This predictor is depicted in Figure~\ref{fig:tage}, borrowed from Seznec and Michaud \cite{seznec2006case}. To predict a branch direction, the PC of the branch is combined with the global branch history to form indices to the different tables, with the index to each tagged using different lengths of the global branch history, with the length growing geometrically. The prediction is provided by the component that uses the longest history and hits (the \textit{provider} component). If there are no hits in the tagged component, the bimodal table provides the prediction. On a misprediction, an entry is allocated in a table that uses a longer history than the \textit{provider}, either the immediate next table, or a random table. It is also possible to allocate multiple entries every time \cite{seznec201164}. TAGE also uses the notion of \textit{alternate} component, which is the table that hits and uses the second most longest history, i.e., it would have been the \textit{provider} if the \textit{provider} had missed.

\begin{figure}[h]
\centerline{\includegraphics[width=\linewidth]{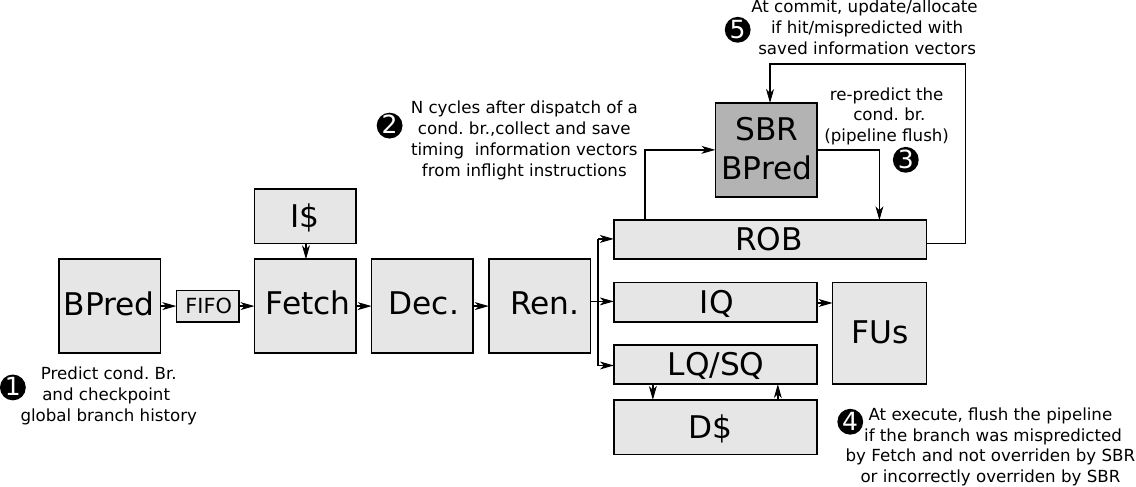}}
%AP: Making space
% SBR operates post-alloc and overrides predictions from the fetch predictor using timing information gathered on the fly.
\caption{SBR (darker) and baseline (lighter) pipeline.}
\label{fig:pipeline_sbr}
\end{figure}

\begin{figure}
\centerline{\includegraphics[trim={0cm 0cm 0cm 0cm},clip,width=\linewidth]{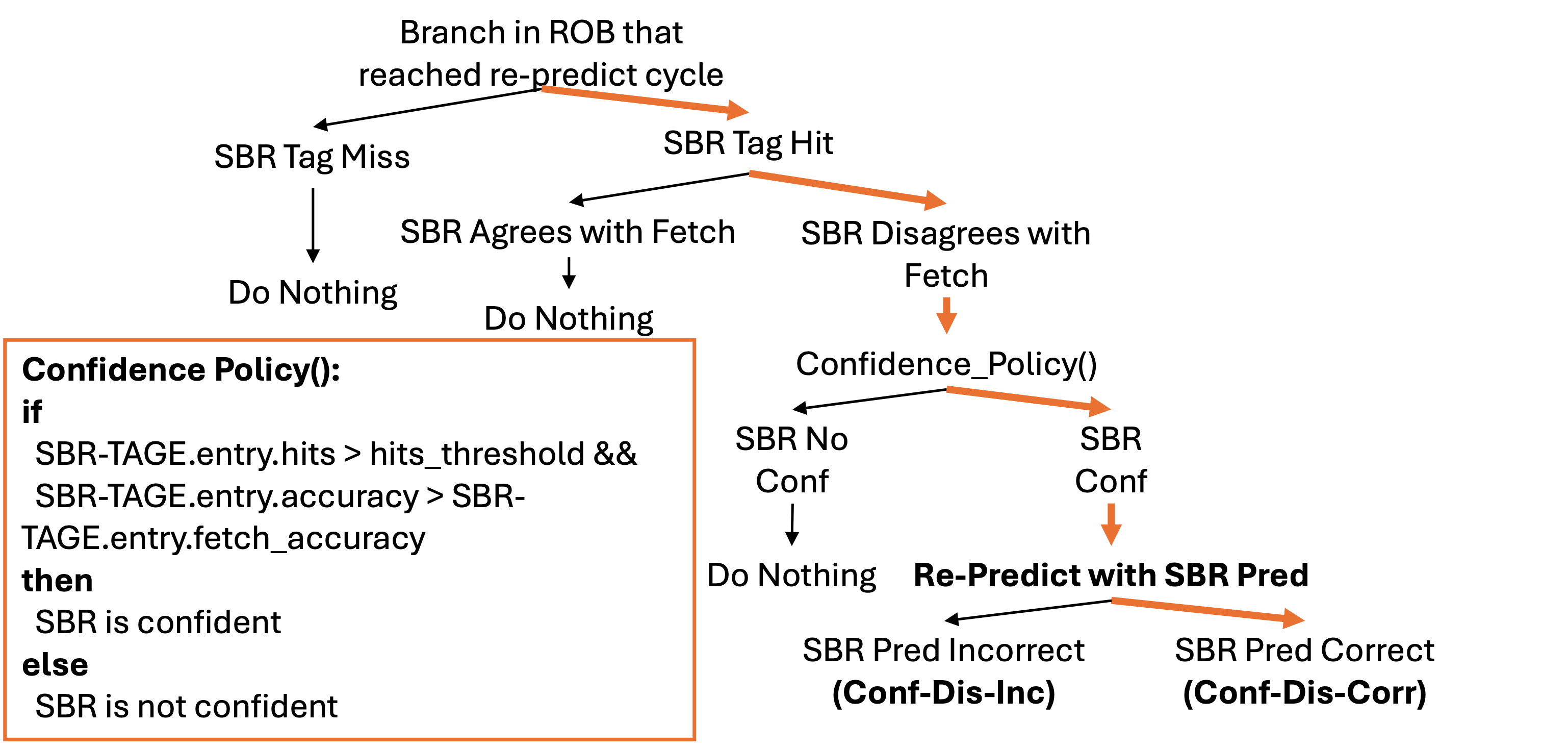}}
\caption{SBR TAGE override decision tree.}
\label{fig:sbr_cases}
\end{figure}

\paragraph{SBR Overview} 
%AP: 3/12 - Removing
%As previously mentioned, the \textit{re-predict} cycle, $n$, is the number of cycles SBR waits after a branch allocates into the ROB before re-predicting it. 
At the \textit{re-predict} cycle, $n$, \textit{SBR TAGE} is accessed using an information vector composed of the CTIV (recently committed branches), OTIV (older branches in the ROB) and YTIV (younger branches in the ROB) hashed with the branch PC. If i) \textit{SBR TAGE} hits, ii) The prediction is different from the prediction of \textit{Fetch TAGE}, and iii) The SBR prediction is confident, SBR overrides the \textit{Fetch TAGE} prediction. If \textit{SBR TAGE} misses then the branch is not re-predicted, but an entry is allocated in \textit{SBR TAGE} if \textit{Fetch TAGE} turns out to be incorrect. For update, we use the same policies as a normal TAGE predictor. Allocation takes place only when i) \textit{Fetch TAGE} mispredicted and ii) \textit{Fetch TAGE} accuracy on the entry SBR had a hit on has accuracy lower than a threshold, or when i) \textit{Fetch TAGE} mispredicted and ii) \textit{SBR TAGE} missed. The high level overview of a pipeline featuring SBR is depicted in Figure~\ref{fig:pipeline_sbr}.

%\todo[inline]{Do we really use the same allocation policy? Or do we allocate only if Fetch TAGE mispredicted (regardless of if we hit or miss in SBR TAGE?}

Furthermore, Figure~\ref{fig:sbr_cases} details the different cases in which SBR either does override fetch (orange arrows) or not (dark arrows). Note that SBR makes use of a heuristic to determine if the prediction of \textit{SBR TAGE} should be preferred over \textit{Fetch TAGE}. The confidence heuristic is listed in Figure~\ref{fig:sbr_cases} and considers counters about the accuracy of \textit{Fetch TAGE} and \textit{SBR TAGE}. Practically, \textit{SBR TAGE} overrides \textit{Fetch TAGE} in a single case, which is when \textit{SBR TAGE} hits, disagrees with \textit{Fetch TAGE}, and is confident according to the heuristic. In all other cases, \textit{SBR TAGE} does not override, that is i) When it misses, ii) When it agrees with \textit{Fetch TAGE}, or iii) When it is not confident according to the heuristic. Table \ref{tab:sbr_categ} summarizes the different possibilities noting for each case that SBR can be correct or incorrect. Once SBR determines that it must override a branch, the pipeline is flushed and the frontend is resteered to the other path of the conditional branch. 

%\todo[inline]{Arthur: I do not like that a lot, but I think it is interesting to have as we talk about Figure 7. Maybe there is a way to make them appear on Figure 7. I think they should match whatever is on Figure 11.}

\begin{table}
%\label{table:sbr_categ}
%AP Making space
%Sub-categories can be combined to form a complete category e.g., Conf-Agree-Corr.
\caption{Predictions categories. Override cases are in \textcolor{orange}{\textbf{bold orange}}.}
\centering
\scalebox{0.80}{
\begin{tabular}{|cccc|}
\hline
\multicolumn{4}{|c|}{SBR TAGE Hit}                                                                                                                                                                     \\ \hline
\multicolumn{2}{|c|}{\textbf{Conf}ident}                                                                                      & \multicolumn{2}{c|}{\textbf{NotConf}ident}                                     \\ \hline
\multicolumn{1}{|c|}{}                                      & \multicolumn{1}{c|}{\textbf{Corr}ect}                           & \multicolumn{1}{c|}{}                                      & \textbf{Corr}ect   \\ \cline{2-2} \cline{4-4} 
\multicolumn{1}{|c|}{\multirow{-2}{*}{\textbf{Agree} with Fetch}}    & \multicolumn{1}{c|}{\textbf{Inc}orrect}                         & \multicolumn{1}{c|}{\multirow{-2}{*}{\textbf{Agree} with Fetch}}    & \textbf{Inc}orrect \\ \hline
\multicolumn{1}{|c|}{}                                      & \multicolumn{1}{c|}{\textcolor{orange}{\textbf{Correct}}}   & \multicolumn{1}{c|}{}                                      & Correct  \\ \cline{2-2} \cline{4-4} 
\multicolumn{1}{|c|}{\multirow{-2}{*}{\textbf{Disa}gree with Fetch}} & \multicolumn{1}{c|}{\textcolor{orange}{\textbf{Incorrect}}} & \multicolumn{1}{c|}{\multirow{-2}{*}{\textbf{Disa}gree with Fetch}} & \textbf{Inc}orrect \\ \hline
\end{tabular}
}
\label{tab:sbr_categ}
\end{table}

This is a double edged sword: If the branch was originally mispredicted by fetch and SBR correctly re-predicted it, then the branch resolution latency has been improved by X cycles, where X is the remaining cycles this branch would need to resolve without SBR. On the contrary, if the branch was originally correctly predicted by fetch and we re-predict it, latency increases by Y cycles, where Y is the pipeline refill penalty after a branch misprediction. As a result, although SBR has potential for saving many cycles it should be used judiciously.
%if slow-to-resolve H2P branches can be re-predicted fast, its application should be parsimonious.

%AP: 3/12 - Removing
%as the penalty of an SBR misprediction can be large for H2P branches that resolve fast, hence the addition of a confidence heuristic.

%A statistic we are interest about in our work is the number of mispredicts we saved, which can be calculated with the SBR Corrects -- SBR Incorrects.

\section{Experimental Setup} \label{exp_setup}

% Table caption is usually above, Figure below, don't ask me why.
\begin{table}[ht]
    \centering
    \caption{Description of benchmarks and H2P branches}
    \scalebox{0.80}{
      \begin{tabular}{|c|c|c|c|c|c|}
      \hline
        Benchmark & Total BMPKI & SC BMPKI & \# of H2P & BMPKI H2P & Coverage \\
        \hline
        \hline
        Blender & 1.71 & 0.24 & 8 & 1.25 & 73\% \\
        \hline
        Deepsjeng & 3.37 & 0.53 & 10 & 1.43 & 42\% \\
        \hline
        Exchange2 & 0.74 & 0.12 & 3 & 0.23 & 30\% \\
        \hline
        Gcc & 3.83 & 0.59 & 16 & 0.37 & 10\% \\
        \hline
        Leela & 8.68 & 1.14 & 17 & 5.54 & 64\% \\
        \hline
        Mcf & 19.93 & 3.34 & 9 & 17.84 & 90\% \\
        \hline
        Omnetpp & 3.22 & 0.61 & 3 & 2.51 & 78\% \\
        \hline
        Wrf & 1.09 & 0.13 & 5 & 0.36 & 33\% \\
        \hline
        X264 & 1.23 & 0.20 & 5 & 0.74 & 60\% \\
        \hline
        Xz & 5.34 & 0.77 & 16 & 4.02 & 75\% \\
        \hline
    \end{tabular}
    }
    \label{tab:h2p}
\end{table}

\subsection{SPEC2017 rate Benchmarks Characteristics}

We evaluate SBR using  SPEC2017 rate benchmarks \cite{bucek2018spec} (compiled for x86\_64 with gcc8 -O2). We selected 92 H2P branches from 10 high BMPKI (> 0.70 BMPKI) benchmarks. A H2P branch has more than 0.045 BMPKI and all selected H2P branches of one benchmark should have at least a cumulative misprediction coverage of 30\%. One exception is \textit{gcc}, as this benchmark has many unique instructions that have a high misprediction rate but are executed only a few times. Table \ref{tab:h2p} describes, for each benchmark, the global BMPKI of \textit{Fetch TAGE}, the BMPKI from the SC component of \textit{Fetch TAGE}, the number of H2P branches we selected for each benchmark and the coverage of these H2P branches. We focus on re-predicting \textit{Fetch TAGE} predictions that were provided by the TAGE components (including bimodal), as they represent the majority of the mispredictions. 

%A widespread feature of SPEC2017 benchmarks is that a few branches are responsible for the 90\% of the branch mispredicts. %This is depicted in Figure \ref{fig:static_vs_dynamic}, in which the number of unique branches is on the x-axis (logscale) and the cumulative BMPKI on the y-axis. The only exception is \textit{gcc} in which around 10K static branches are needed to cover 90\% of the branch mispredicts. 

%\todo[inline]{Is it really such a good argument for ignoring SC mispreds? Did we just forget to do it or there is another more fundamental reason?}

%\begin{figure}
%\centerline{\includegraphics[trim={0cm 0cm 0cm 0cm},clip,width=\linewidth]{static_vs_dynamic.png}}
%\caption{BMPKI and Static branches for SPEC2017 Benchmarks}
%\label{fig:static_vs_dynamic}
%\end{figure}

%Fig.~\ref{fig:static_vs_dynamic} depicts the number of static branches (x-axis logscale) that are responsible for the BMPKI (y-axis) of each benchmark. For all the benchmarks except \textit{gcc},  few branches are responsible for 90\% of the mispredicts. gcc (light blue line) shows a different behaviour as to cover 90\% of the mispredicts, thousands of static branches have to be considered.

\subsection{Simulation Framework}

We find representative regions of workloads following the Simpoint sampling methodology \cite{hamerly2005simpoint,patil2010pinplay}. We then replay these simpoints in \textit{gem5} \cite{lowe2020gem5}, using full-system mode (50M warm-up, 100M actual run). We model a detailed out-of-order pipeline resembling Intel Skylake (6-wide, 224 ROB, 97 IQ, 72 LQ, 56 SQ). The model features a decoupled frontend that implements Fetch Directed Instruction Prefetching (FDIP) with a 192-entry fetch target queue \cite{reinman1999fetch} and a very large (32M-entry tables) \textit{Fetch TAGE-SC} branch predictor. The latter is used to minimize aliasing effects such that if SBR is able to capture behavior that TAGE-SC cannot, it will be because of a fundamental difference in the information it uses to predict, rather than because \textit{Fetch TAGE-SC} suffers from aliasing. For cache prefetching we use the \textit{tagged prefetcher} as we found it to be the best performing data prefetcher among all those currently available in gem5. The \textit{SBR TAGE} predictor also uses very large tables and the same history length as the \textit{Fetch TAGE}. However, \textit{SBR TAGE} has neither bimodal table nor a SC component. Another difference is that \textit{SBR TAGE} tables are updated/allocated only from the selected H2P branch but the history, as the original TAGE, is updated by all the branches.% Table \ref{table:gem5_config} summarizes the pipeline parameters used for the evaluation of SBR.

\subsection{Design Space Exploration}

We explored all the combinations of the following parameters.
%AP: Making space
%Our analysis explored a rich design space.
%As we illustrate in Section \ref{results}, branches exhibit varying resolution cycles after entering the ROB. As a result, 
We investigate the \textbf{re-predict cycle} used for SBR, to determine the number of branches eligible for SBR as well as the quality of the timing information available at specific re-predict cycles.
%IC: Making space
%
%AP: Making space
%recall that the SBR information vector encodes whether a branch resolved fast or slow. As a result, 
We vary the \textbf{timing thresholds} of the different TIVs that determine whether resolution was fast or slow. We sweep a range from 2 to 128 cycles, in powers of 2. Note that the timing threshold for CTIV and OTIV is the same, but the YTIV threshold is different as its behavior is tied to the re-predict cycle. For example, if the re-predict cycle is 16, there is no reason to have a threshold above 16 for YTIV as all of YTIV bits will encode \textit{fast}. As a result, when YTIV is included we explore the timing thresholds from 2 until the re-predict cycle in powers of 2. 

%This also implies that there is an inconsistency between CTIV/OTIV and YTIV since the same bit of information represents a slightly different quantity.

We consider vector combinations
%To assess the trade-off between history inconsistency and the availability of "future" branch information in the history, 
using only older information (CTIV+OTIV) and using both older and younger information (CTIV+OTIV+YTIV). %Similarly, instead of concatenating vectors (using least significant bits for younger information, most significant bits for older information), we consider XOR-ing them, to increase the weight of older information.
We also consider XOR-ing the overall TIV with the global branch history, as global history is known to be good at finding correlations between branches \cite{seznec2006case,jimenez2001dynamic,mcfarling1993combining,seznec2007ltage,seznec2014tage,smith1981study,falcon2004prophet,zangeneh2020branchnet}.

%AP: Making space

We varied the allocation policy for \textit{SBR TAGE}. Our scheme allocates in \textit{SBR TAGE} on \textit{Fetch TAGE} misprediction, but checks if the \textit{Fetch TAGE} accuracy of an entry is below a threshold. The intuition behind this is that if a \textit{Fetch TAGE} entry provides good accuracy, then there is no point for allocating an \textit{SBR TAGE} entry for it. In our experiments, we considered 2 accuracy thresholds. First, the \textit{Fetch TAGE} entry must have below 99\% accuracy, as to never limit the allocations of SBR and train it quickly. Second, the \textit{Fetch TAGE} entry must have below 80\% accuracy, to filter out cases where \textit{Fetch TAGE} is doing well enough.

\section{Results} \label{results}

\subsection{Branch Resolution Latency}

\begin{figure}[t]
\centerline{\includegraphics[width=0.9\linewidth]{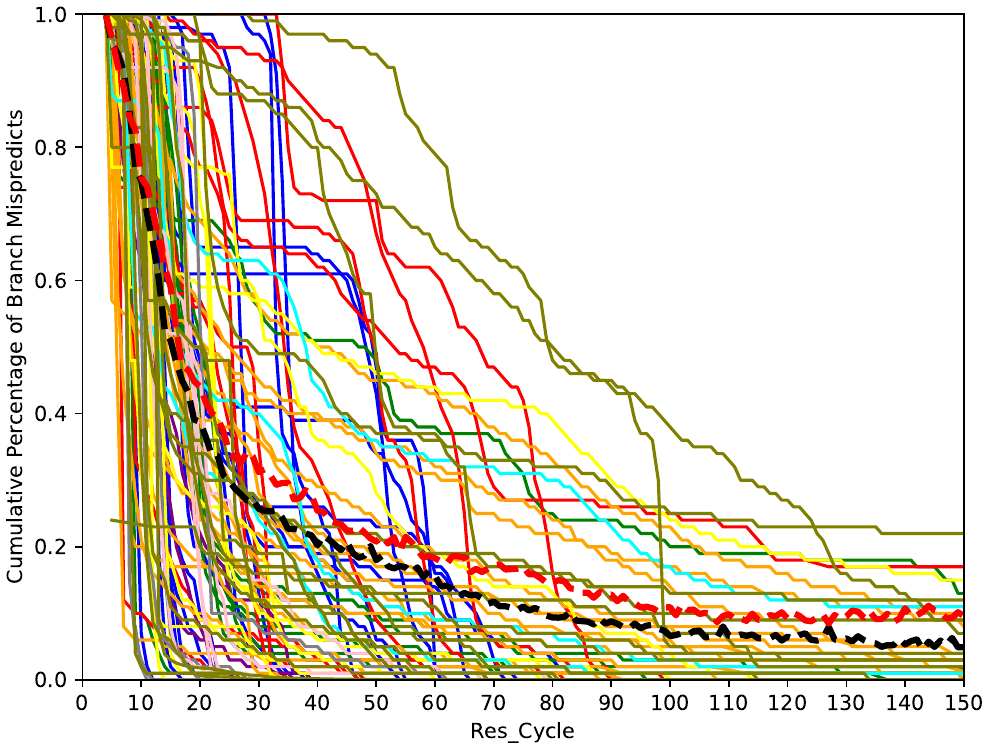}}
\caption{H2P mispredictions resolution cycle (coloured per benchmark) - Black dashed line is Skylake-Like core average - Red dashed line is Goldencove-Like core average}
\label{fig:res_cycle}
\end{figure}

We first show a characterization of the resolution latency (after entering the ROB) of the selected H2P branches when they mispredict. This is an important analysis as the SBR re-predict cycle is a tradeoff between how many H2P branches can be repredicted and how "populated" the timing information vector will be. Moreover, this analysis denotes the cycles we can potentially save for each H2P branch. Figure \ref{fig:res_cycle} depicts the cumulated percentage of resolution latencies of all 92 branches (truncated to 150 cycles), with all branches belonging to a given benchmark using the same color. The Figure also shows the average resolution cycle for H2P mispredictions for the Skylake-like and Goldencove-like cores with the black and red dashed lines respectively. 

%We consider the latter in Section \ref{sec:goldencove} to study the impact of a larger instruction window on SBR.

\begin{figure}[h]
\centerline{\includegraphics[width=\linewidth]{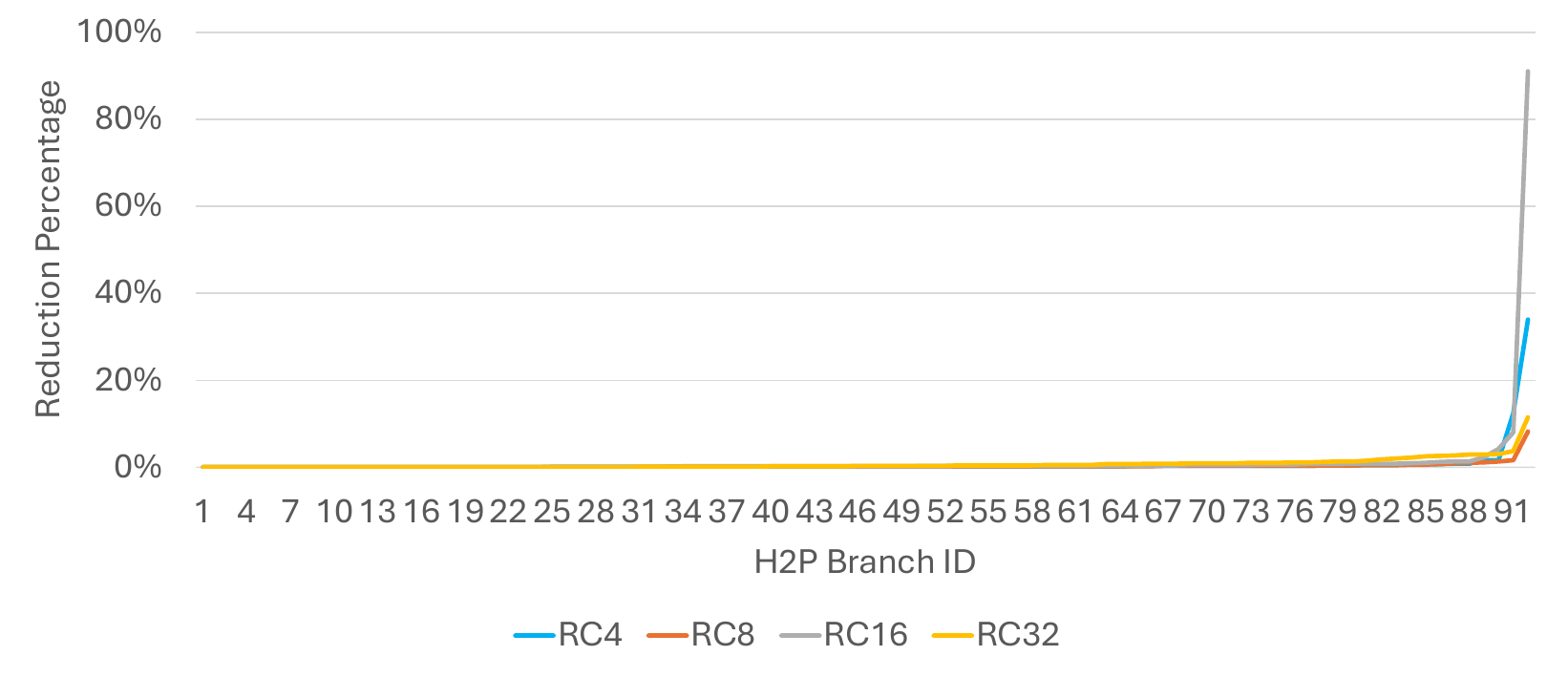}}
\caption{Misp. reduction when varying the re-predict cycle}
\label{fig:gem5_per_rc}
\end{figure}

Consider the uppermost line: Around 40\% of the mispredictions for that branch are resolved within 70 cycles after entering the ROB. This branch therefore has high potential for re-prediction, as we can repredict at cycle 70 and still have the opportunity to catch more than 60\% of the mispredictions. On the other hand, focusing on the bottom left hand corner of the Figure, there are many branches for which 90\% of the mispredictions are resolved within 10 cycles after entering the ROB. Those kind of branches do not provide much headroom for collecting timing information and re-predicting them. By considering the average, which is the dashed black line, it shows that 60\% of the branch mispredicts are resolved within 20 cycles, while the red dashed line which is the Goldencove-like core shows that H2P branch mispredicts tend to take longer to resolve. In terms of saving cycles, based on the average we can save up to 20 cycles for the 40\% of the branch mispredicts and up to 40 cycles for 20\% of the branch mispredicts. Nevertheless, the figure also illustrates that each H2P branch has different behavior in term of resolution cycle, even within the same benchmark. Consider the blue lines, which belong to \textit{leela}, we can observe both behaviors (fast and slow to resolve). This suggests that ideally, each PC would have to be re-predicted at a different time, even within a benchmark.

\begin{figure}[t]
\centerline{\includegraphics[width=\linewidth]{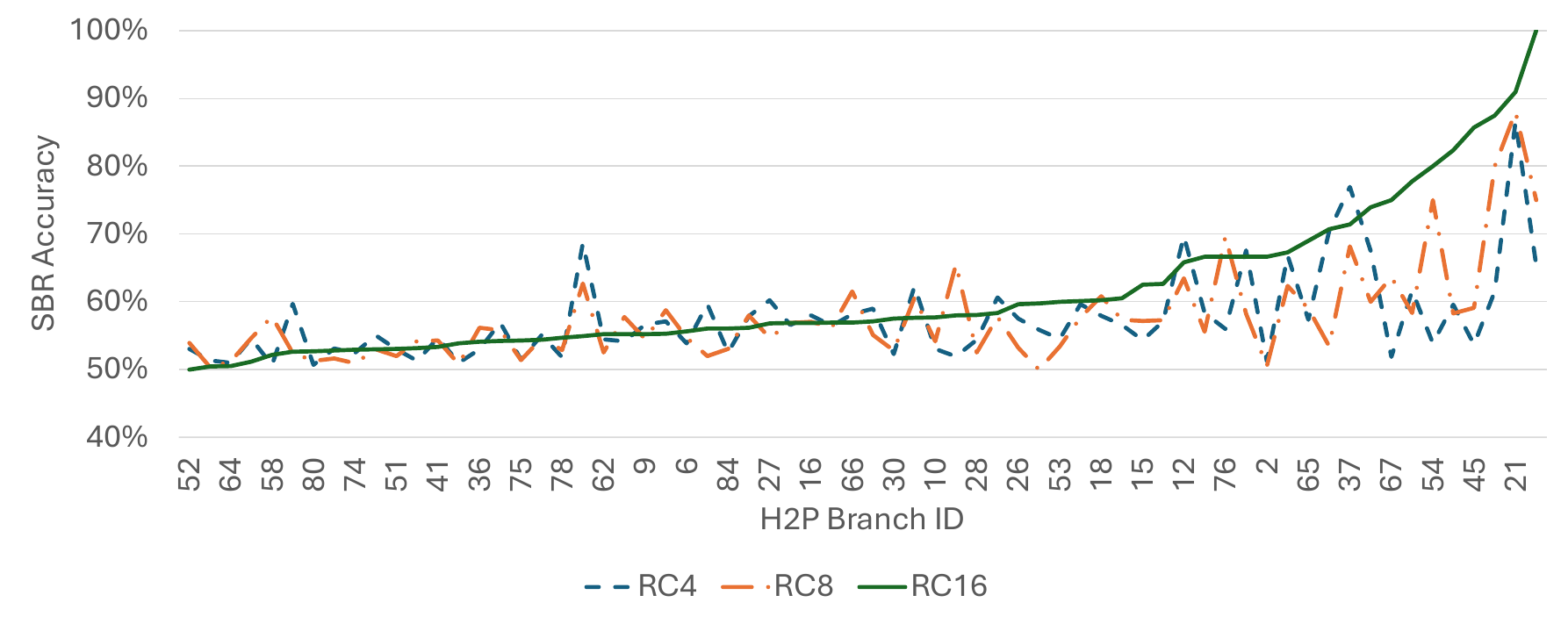}}
\caption{SBR accuracy with different re-predict cycles}
\label{fig:sbr_acc_rc}
\end{figure}

\subsection{Analysis of SBR Potential}

In this work, we aim to gauge the potential of SBR and therefore focus on performance analysis of an idealistic SBR mechanism on top of a baseline out-of-order pipeline, using cycle-level simulation.

\subsubsection{Impact of re-predict Cycle}

%\todo[inline]{@Ioannis: Should we mention the numbers in the Figure are with the best SBR configs?}

Figure \ref{fig:gem5_per_rc} shows the misprediction reduction percentage for each re-predict cycle, for all the H2P branches, using the best configuration in the design space for each branch. The four lines of this figure are sorted separately, hence, the x-axis labels do not represent each a unique H2P branch: the same point across the lines can be a different H2P branch. The criterion for picking the best configuration is the largest misprediction reduction. We note that the H2P branch discussed in Section~\ref{motivation} sees a reduction of 91\% at re-predict cycle 16 (\textit{Fetch TAGE} misprediction rate is 82\%). We also observe two cases where the reduction is 33\% and 12\% at re-predict cycle 4. The former is again for the same H2P branch. Nevertheless, the vast majority of the H2P branches do not benefit from SBR.
%in terms of reverting mispredicted branches.
%The labels of the x-axis correspond to the ID of different H2P branches 1--92 (all figures in this Section use same ID to identify each branch).
%however, in this case, SBR does not leverage the timing of the H2P branch itself but the timing of the older branches as described in Section \ref{motivation}.

Figure \ref{fig:sbr_acc_rc} shows the \textit{SBR TAGE} accuracy (y-axis) as a function of the re-predict cycle. Re-predict cycle 32 is not shown as many H2P branches have resolved by that time, resulting in low potential overall. 
%AP Making space
%
 %The labels of the x-axis correspond to the IDs of the different H2P branches. 
 %For example, point 21 relates to the branch discussed in Section \ref{motivation} in both cases. 
 In general, we observe that waiting longer is beneficial, as the SBR accuracy tends to be highest with re-predict cycle 16.
From Figure \ref{fig:sbr_acc_rc} onward, all figures with x-axis labeled as \textit{H2P branch ID} are consistent, e.g., \textit{H2P Branch ID} 21 denotes the H2P branch discussed in Section 3.

% \todo[inline]{@Ioannis: If you have this number, then put how much potential of repredicting we lose by going from RC4 to RC8 and RC16. otherwise, remove this.}

\subsubsection{Performance of Best per-PC SBR Configurations}

%\todo[inline]{Here I think we are missing a high level refresher on what are the possible configs, like mixing TIVs together, encoding, length, etc. Maybe \textbf{how many} configurations you cna have for each PC (and you pick the best).}

\begin{figure*}[t]
\centerline{\includegraphics[width=\linewidth]{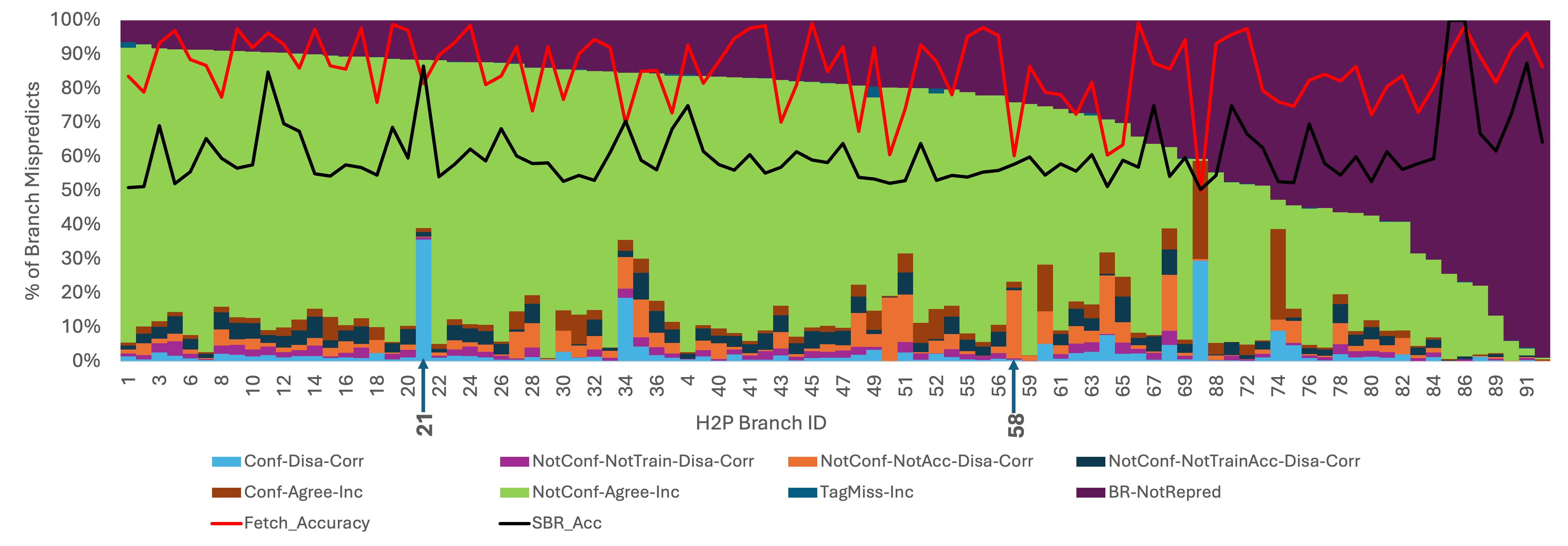}}
\caption{Results of gem5 simulations (Point 21 is mcf branch A introduced in Section 3)}
\label{fig:gem5_res}
\end{figure*}

Figure \ref{fig:gem5_res} reports the contribution of different classes of mispredictions for each of the 92 PCs, using the best timing vector configuration per PC. The 8 categories are the ones described in Table \ref{tab:sbr_categ}. We also show two lines which represent the \textit{Fetch TAGE} accuracy (red line) and the \textit{SBR TAGE} accuracy (black line). %There are 6 classes relating to SBR TAGE hitting, based on whether i) SBR TAGE was confident or not confident according to the confidence heuristic ii) SBR TAGE agrees or disagrees with Fetch TAGE and iii) SBR TAGE was actually correct. The lack of confidence of SBR TAGE can then be attributed to the provided entry being not yet trained, not accurate, or both. The two remaining classes are SBR TAGE misses (M-I) and branches that resolve before the re-predict cycle (BR-NR).

The largest category corresponds to the case where \textit{Fetch TAGE} was incorrect, \textit{SBR TAGE} agrees with \textit{Fetch TAGE} and \textit{SBR TAGE} is not confident (NotConf-Agree-Inc). This suggests that either the mechanism or the predictor we use is not suitable for this kind of information, or the information we use fails to differentiate the mispredictions from the correct predictions. Another frequent case is BR-NotRepred, in which branch instances do not reach the re-predict cycle, confirming that many H2P branches often resolve fast (Figure \ref{fig:res_cycle}). SBR is beneficial only in the \textit{Conf-Disa-Corr} category, which is limited and only covers more than 10\% of the \textit{Fetch TAGE} mispredictions for 3 branches. 

%Note that this does not necessarily mean that SBR is beneficial as the Figure does not show \textbf{how many} predictions were incorrectly reversed by SBR.

It is interesting to note that 2 "winner" branches share the same best configuration, which uses COTIV only, meaning older branches, at re-predict cycle 4 and timing threshold 128 cycles. This is not unexpected as those two branches are in the same function, \textit{spec\_qsort} (one is branch A, number 21), suggesting that both H2P branches share the same timing behavior, and that SBR is able to capture it. We also analysed some of the non-beneficial H2P branches best configurations and most of them use the configuration that XORs the global branch history with the TIV. This suggests that for those branches, timing does not correlate with branch directions, and adding the ghist simply allows \textit{SBR TAGE} to reach the same conclusions as \textit{Fetch TAGE}.

\begin{figure}[t]
\centerline{\includegraphics[trim={0cm 0cm 0cm 0cm},clip,width=\linewidth]{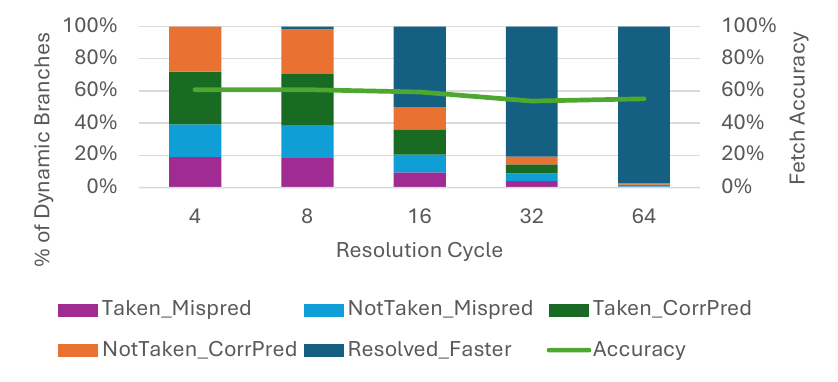}}
\caption{Taken and Not-Taken Correct Predictions and Mispredictions for an omnetpp branch as a function of resolution time}
\label{fig:omnetpp_mr_rates}
\end{figure}

%\begin{figure}[ht]
%\centerline{\includegraphics[trim={0cm 0cm 0cm 0cm},clip,width=\linewidth]{omnetpp_stack.png}}
%\caption{Stacked column (Like Figure \ref{fig:gem5_res}) for omnetpp Branch Best Configuration}
%\label{fig:omnetpp_stack}
%\end{figure}

\subsubsection{Why SBR is not working}

We explain the limitations of timing information using a H2P branch from omnetpp for which SBR offers no advantage. Listing \ref{list:omnetpp_code} shows the code of the function shiftup and of the inline operator “<=”. The shiftup restructures the heap, denoted by table $h[]$ in the listing, and uses the inline operator for deciding the order relation between heap elements. The elements are cMessages which can represent events, messages, jobs or other entities in a network simulation. We analyzed the branch in Line 22. This branch is part of a series of checks that first compare the timestamps (Lines 19,20) to determine if the messages occurred at different times. If the timestamps are identical, the branch compares the scheduling priorities of the messages (Lines 21,22), which are derived from the input cMessages. These priorities are a function of the cMessages and the specific heap configuration. This means that the branch outcome is determined by the data characteristics of messages, i.e, is data dependent. First, Figure \ref{fig:omnetpp_mr_rates} establishes the lack of self timing correlation of the branch in Line 22, by showing that its Fetch TAGE accuracy is insensitive to its resolution cycle. 

\begin{lstlisting}[language=C, caption=Source code of shiftup in omnetpp,label={list:omnetpp_code},basicstyle=\fontsize{6}{6}\selectfont\ttfamily]
void cMessageHeap::shiftup(int from){
    // restores heap structure (in a sub-heap)
    int i = from,j;
    cMessage *temp;
    while ((j=i<<1) <= n){
        if (j<n && (*h[j] > *h[j+1]))   //direction
            j++;
        if (*h[i] > *h[j]){  //is change necessary?
            temp=h[j];
            (h[j]=h[i])->heapindex=j;
            (h[i]=temp)->heapindex=i;
            i=j;
        }
        else
            break;
    }
}
inline int operator <= (cMessage& a, cMessage& b){
    return (a.getArrivalTime() < b.getArrivalTime()) ? 1 :
    (a.getArrivalTime() > b.getArrivalTime()) ? 0 :
    (a.getSchedulingPriority() < b.getSchedulingPriority()) ? 1 :
    (a.getSchedulingPriority() > b.getSchedulingPriority()) ? 0 :
    a.getInsertOrder() <= b.getInsertOrder();
}
\end{lstlisting}

%which denotes data dependency., rather than timing correlation. <== i think this is unbased

%\textcolor{red}{To assess that in general, we analysed the correlation between past branches and the actual direction of branch A. We created a vector of length 10 with setting a bit to 1 if a past branch required less than 20 cycles to resolve and to 0 otherwise. We experimented with various lengths and observed that the overall trends remain the same. For each unique vector we counted the number of instances where branch A was actually taken and not-taken and then we plotted these counts on a scatter plot. Figure \ref{fig:vectors_scatter_mcf} depicts the results of this analysis and it reveals a clear bias towards the not-taken direction. This indicates that other branches timing information are more frequently associated with not-taken instances than with taken.}

%This establishes that the branch itself does not exhibit any timing correlation.

Second, to investigate why timing correlation with other branches is not useful for this branch, we analyzed the correlation of the direction of this branch with its TIV vectors. We used TIV vectors that include the ten most recent branches from COTIV and YTIV at repredict cycle 8 (we checked various vector lengths, but the conclusions remained the same). Older branches include those at Lines 19–21 (always executed before Line 22), while younger branches include Line 23 and branches from subsequent iterations. For each unique vector we count how often the branch is taken and not-taken to determine a vector’s SBR bias, defined as the most popular direction divided by the total vector frequency.

\begin{figure}
\centerline{\includegraphics[trim={0cm 0cm 0cm 0cm},clip,width=\linewidth]{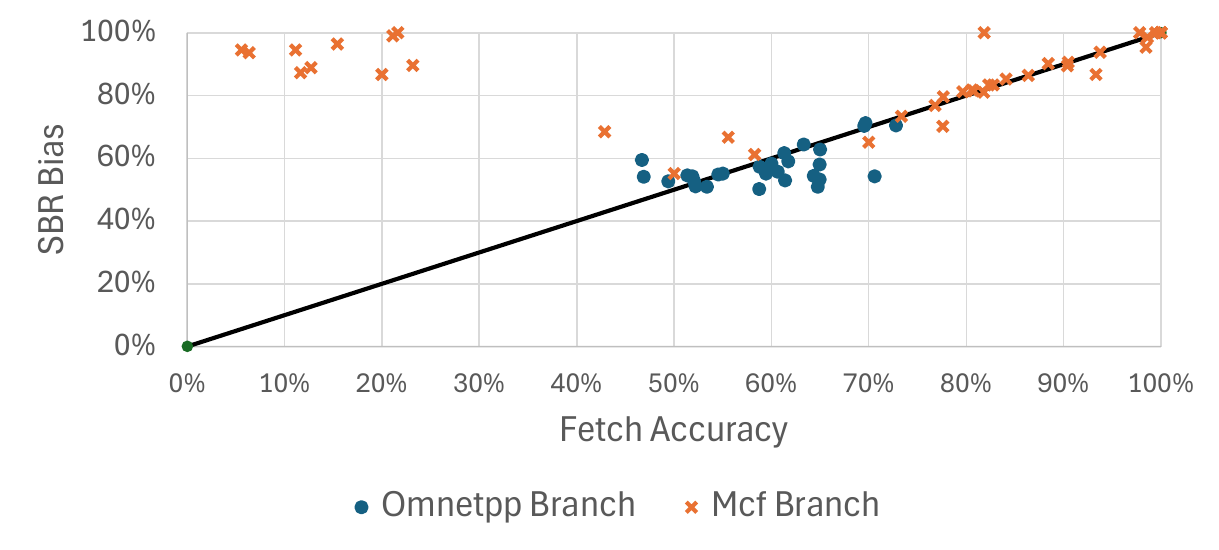}}
\caption{Scatter plot of vectors encoded with 1 and 0 based on resolution cycles for omnetpp H2P branch and mcf H2P branch - The black line is a reference of the diagonal}
\label{fig:vectors_scatter_analysis}
\end{figure}

The scatter plot in Figure \ref{fig:vectors_scatter_analysis} compares the SBR bias against the \textit{Fetch TAGE} accuracy for each vector. Points along the diagonal (black line) indicate that SBR TIVs do not glean any additional correlation. We observe that for the \textit{omnetpp} branch, all vectors align with the diagonal, confirming the lack of additional correlation from the use of timing information as compared to \textit{Fetch TAGE}. 

For the \textit{H2P Branch ID 58} (Branch at Line 22), Figure \ref{fig:gem5_res} shows that the portion of mispredicts where SBR agrees with the fetch predictor (green segment) dominates, along with the portion where SBR disagrees but it's not confident enough to re-predict (orange segment). This confirms that timing information does not help capture better the correlation behavior of this branch as compared to \textit{Fetch TAGE}.

We also perform the bias analysis, for the H2P branch in \textit{mcf} (Section \ref{motivation}), using only older branches at repredict cycle 4. Younger branches do not improve the SBR accuracy for this branch for any SBR configuration we evaluated. In contrast to the branch from \textit{omnetpp}, Figure \ref{fig:vectors_scatter_analysis} reveals that \textit{mcf} has vectors well above the diagonal, indicating cases where SBR can improve over Fetch TAGE by leveraging timing information.

\subsubsection{Goldencove Analysis}
\label{sec:goldencove}

The temporal behavior of instructions in the pipeline depends on the microarchitecture. Hence, we study the impact of larger structures on SBR using a Goldencove-Like core. This core is 8-wide with 512 ROB entries, 220 IQ entries, 172 LQ and 156 SQ entries, 350 integer and 330 floating-point registers. 
%We depart from the actual Golden Cove microarchitecture by implementing 64KB 16-way associative L1-I and L1-D caches, a 2MB L2 and a 8MB L3. All caches have 32 MSHRs. All the other parameters are the same as the Skylake-Like core.

We explored the same design space in terms of \textit{SBR TAGE} configurations, and we observed (results not shown due to space limitations) that with a larger window, SBR accuracy tends to be slightly higher, although SBR remains unable to provide gains except in few specific cases. However, H2P branches tend to take slightly longer to resolve with a longer instruction window, as seen in Figure~\ref{fig:res_cycle}, suggesting more SBR opportunity with a larger window in terms of how many instances can be re-predicted before a given re-predict cycle.

%\todo[inline]{@Ioanni: What we discussed}.

%\subsection{Why is Potential so Low?}

%\todo[inline]{Would be good to have a discussion about what is happening? Maybe some branches depend on random? Depend on unpredictable data? etc.}

%\subsection{Analysis of the Winners}
%\todo[inline]{Analyse winners vectors, short/long history tables}
%\todo[inline]{Arthur: This will be done with Figure 12 discussion (winner vector info) + TAGE provider in 6.2.3}

%\todo[inline]{At least show reslution cycle of H2P branches, if time permits show results on exploration or subset of exploration}
\section{Conclusion}
This paper set out to explore if using new information that is available later in the pipeline -- timing information -- can help re-predict hard-to-predict (H2P) branches with better accuracy. The main conclusion is that the information we considered, as leveraged by a TAGE predictor, cannot beat a large TAGE-SC predictor that only uses architectural information except in very specific cases.

Although the result is generally negative, i.e., timing as defined in this work rarely correlates with higher accuracy to branch outcome as compared to a large TAGE-SC, we identified two H2P branches that benefit from timing information, and provided an in-depth analysis as to why the timing information is useful for one of them. As a result, this study can serve as a starting point for SBR-like mechanisms that use a different kind of post-fetch microarchitectural information. We have also observed that while some H2P branches usually take a long time to resolve, the average resolution time (allocation in ROB to execution) is below 20 cycles for 50\% of the H2P branches studied in this work.
%AP Making space
%of the H2P branches studied in this work.
As a result, future work may consider using microarchitectural information available earlier in the pipeline to refine existing overriding predictors \cite{jimenez2000impact}.

%but our findings from the offline analysis suggest that timing information can have the same accuracy as TAGE-SC with shorter history.

%For Future work, we will re-predict the H2P mispredictions that are provided from the SC...try different kind of information...

%\todo[inline]{Need to work on it...}

\section*{Acknowledgments}

This work was partially supported by an Intel Academic Grant. The authors would like to thank Jared Stark and Jayesh Gaur for their guidance and support.

\bibliographystyle{IEEEtranS}
\bibliography{IEEEexample}

\end{document}